\documentclass[prd,aps,twocolumn,preprintnumbers, showpacs, nofootinbib,notitlepage]{revtex4-1} 
\usepackage{amssymb,amsthm,amsmath} 
\usepackage{multirow}
\usepackage{makecell}
\usepackage{float}
\usepackage{url}
\usepackage{xcolor}
\usepackage{booktabs}
\usepackage[T1]{fontenc} 
\usepackage{feynmp}
\usepackage{graphicx}
\begin{document} 
\title{\boldmath Puzzles in charmed baryon semileptonic decays with $SU(3)_F$ flavor symmetry and lattice inputs
}

\author{Chao-Qiang Geng$^1$, Chia-Wei Liu$^1$, Sheng-Lin Liu$^{1,2,3}$}\email{liushenglin22@mails.ucas.ac.cn} 	
\affiliation{$^1$School of Fundamental Physics and Mathematical Sciences, Hangzhou Institute for Advanced Study, UCAS, Hangzhou 310024, China}
\affiliation{$^2$Institute of Theoretical Physics,~UCAS, Beijing 100190, China}
\affiliation{$^3$University of Chinese Academy of Sciences, 100190 Beijing, China}

\begin{abstract}
	Recent measurements of charmed-baryon semileptonic decays signal large
		tensions between experiment and theory, including $SU(3)_F$ analyses and
		lattice-quantum chromodynamics simulations. Possible sources of the discrepancy include the
		experimental normalization mode $\Xi_c^0 \to \Xi^- \pi ^+$. Using
		lattice-QCD inputs in an $SU(3)_F$ analysis including first-order symmetry
		breaking, we predict
		$ 
			{\cal B}( \Xi_c^+ \to \Sigma^0 \ell^+ \nu_\ell )
 / 
			{\cal B}( \Xi_c^+ \to \Xi^0 \ell^+ \nu_\ell )
		=(2.6\pm0.3)\%,
		$
		and
		$ 
			{\cal B}( \Xi_c^+ \to \Lambda \ell^+ \nu_\ell )
/ 			{\cal B}( \Xi_c^+ \to \Xi^0 \ell^+ \nu_\ell )
		=(1.1\pm0.1)\%,
		$ with $\ell^+=(e^+,\mu^+)$.
		These ratios provide normalization-independent tests 
		and may help clarify the origin of the present tension. 
\end{abstract}

\maketitle
\flushbottom

\section{Introduction}
\label{sec:intro}
Significant discrepancies in $\mathcal{B}(\Xi_c^0\rightarrow\Xi^- e^+\nu_{e})$ have been found between the experimental data and theoretical studies, including lattice-quantum chromodynamics~(LQCD). The relative branching fraction with respect to $\Xi_c^0 \to \Xi^- \pi^+$ was measured to be $0.730\pm0.044$~\cite{Belle:2021crz} and $0.825\pm0.124$~\cite{ALICE:2021bli,ALICE:2025wrq} at Belle and ALICE, respectively. Using the reference channel $\mathcal{B}(\Xi_c^0\rightarrow\Xi^-\pi^+)=(1.43 \pm 0.27)\%$~\cite{ParticleDataGroup:2024cfk}, the corresponding absolute branching fractions are $(1.04\pm0.21)\%$ and $(1.18\pm0.28)\%$. In contrast, LQCD predicts $(2.48\pm0.86)\%$~\cite{Zhang:2021oja} and $(3.58\pm0.12)\%$~\cite{Farrell:2025gis}. On the other hand, $SU(3)_F$-based analyses predict that the absolute branching fraction should be around $4\%$, using $\Lambda_c^+ \to \Lambda e^+ \nu_e$ measured at BESIII as input~\cite{He:2021qnc}. The $SU(3)_F$ prediction is in strong tension with the data, but in reasonably good agreement with the LQCD results. The experimental data indicate that the $SU(3)_F$ breaking exceeds $50\%$, and a theoretical understanding of this issue has remained elusive for more than half a decade. 

Since both Belle and ALICE collaborations have reported convergent results, one naturally suspects that the discrepancy arises from an underestimation of the normalization channel $\mathcal{B}(\Xi_c^0 \to \Xi^- \pi^+)$~\cite{Geng:2023pkr,Cheng:2024lsn}. 
A determination of this branching fraction is crucial, as it is so far the only absolute branching fraction that has been measured for $\Xi_c^0$, and its uncertainty propagates into other measurements. 
Recently, a similar underestimation of the experimental branching fraction ${\cal B}(\Xi_c^+ \to \Xi^- \pi^+ \pi^+)$ has also been conjectured~\cite{Geng:2024sgq,Yang:2025orn}. 
In particular, the $SU(3)_F$ framework incorporating the K\"orner-Pati-Woo theorem suggests that this channel may be underestimated by a factor of two~\cite{Geng:2023pkr,Cheng:2024lsn}, with a discrepancy of about $4\sigma$ from the experimental result, although a fully satisfactory explanation remains under debate~\cite{Aliev:2021wat,Wang:2025khg,Aliev:2025zbk}.

In this work, 
we parameterize the form factors of singly charmed baryon decays using $SU(3)_F$ flavor symmetry in the Bourrely-Caprini-Lellouch (BCL) $z$-expansion, including the first-order symmetry-breaking effects for the first time. The corresponding parameters are then determined from the LQCD results for these decays. The approximate symmetry among the light quarks provides a model-independent framework that has been widely applied to heavy-hadron decays~\cite{Savage:1989qr,Savage:1991wu,Pirtskhalava:2011va,Grossman:2012ry,Lu:2016ogy,Geng:2017esc,Geng:2017mxn,Wang:2017gxe,Geng:2018plk,Geng:2018bow,Geng:2018rse,Geng:2018upx,Cen:2019ims,Hsiao:2019yur,Geng:2019awr,Geng:2019bfz,Jia:2019zxi,Geng:2019xbo,He:2021qnc,Liu:2023dvg,Hsiao:2023mud,Geng:2023pkr,Zhong:2024zme,Geng:2024sgq,Zhong:2024qqs,He:2024unv,Sun:2024mmk,Xing:2024nvg,Cheng:2024lsn,Geng:2025wfi,Wu:2025hnh,Cheng:2025oyr,Yang:2025orn}. We further propose golden channels that can help clarify the discrepancies observed experimentally. 

This paper is organized as follows. 
In Sec.~\ref{sec:Formalism}, we present the formalism of the $SU(3)_F$ flavor-symmetry approach including first-order breaking. 
In Sec.~\ref{sec:Numerical results}, we provide numerical results for the form factors, branching fractions, and angular observables of each channel, followed by discussions. Our conclusions are given in Sec.~\ref{sec:Conclusions}.

\section{Formalism}
\label{sec:Formalism}   

The transitions of  ${\bf B}_c\to {\bf B}\ell^+\nu_{\ell}$  induced by a current are  best described by form factors, where $\ell^+=(e^+,\mu^+)$ denotes the antileptons, and ${\bf B}_c $ and ${\bf B}$ stand for the charmed baryon and the octet baryon, respectively. 
In this work, we adopt the standard helicity form factors and parameterize their
$q^2$ dependence with the BCL $z$-expansion~\cite{Bourrely:2008za}, defined by 
\begin{eqnarray}\label{eq1}
	&& f \left(q^{2}\right)=\frac{1}{1-q^{2} /\left(m_{\mathrm{pole}}^{f }\right)^{2}} \sum_{n=0}^{n_{\max }} a_{n}^{f }\left[z_{f }\left(q^{2}\right)\right]^{n},
    \nonumber\\
&&z_{f }\left(q^{2}\right)=\frac{\sqrt{t_{+}^{f }-q^{2}}-\sqrt{t_{+}^{f }-t_{0}}}{\sqrt{t_{+}^{f }-q^{2}}+\sqrt{t_{+}^{f }-t_{0}}}.\\
&&f = f_{+}, f_{\perp},f_0, g_{+}, g_{\perp},g_0, \nonumber
\end{eqnarray} 
with
$q =   p_{{\bf B}_c}-p_{{\bf B}} $ and 
$t_{0}= (m_{{\bf B}_c}-m_{{\bf B}})^{2}$.
For $f= f_{+}, f_{\perp},f_0$, 
the branch point 
$t_{+}^{f} = ( m_D + m_{\pi(K)}) ^2 $ is taken as the threshold of the $D\pi(K)$ two-particle states for the $c\to d(s)$ transitions, respectively. For $f=g _{+}, g_{\perp},g_0$, the corresponding branch point is set by the $D^{*}\pi$ or $D^{*}K$ thresholds, $t_+^f =( m_{D^*} + m_{\pi(K)}) ^2 $.  
To 
 satisfy the end-point relations of $f_0(0)=f_+(0)$ and $g_0(0)=g_+(0)$, we choose
\begin{equation}\label{match}
\begin{aligned}
    \!\!\!\!&a_2^{f_0}\!=\!\frac{a_0^{f_+}-a_0^{f_0}}{z_{f_0}(0)^2}+\frac{a_1^{f_+}-a_1^{f_0}}{z_{f_0}(0)}+a_2^{f_+}+(a_3^{f_+}-a_3^{f_0})z_{f_0}(0),\\
    \!\!\!\! &a_2^{g_0}\!=\!\frac{a_0^{g_+}-a_0^{g_0}}{z_{g_0}(0)^2}+\frac{a_1^{g_+}-a_1^{g_0}}{z_{g_0}(0)}+a_2^{g_+}+(a_3^{g_+}-a_3^{g_0})z_{g_0}(0), 
\end{aligned}
\end{equation}
which were first used in Ref.~\cite{Farrell:2025gis}.

The $SU(3)_F$ flavor-symmetry framework treats the light quarks $u$, $d$, and $s$ on an equal footing, thereby classifying hadrons into multiplets corresponding to irreducible representations of $SU(3)_F$. The baryons are embedded in the $SU(3)_F$ representations
	\begin{equation}\label{new}
		\begin{aligned}
			&\mathbf{B}_c=(\Xi_c^0,-\Xi_c^+,\Lambda_c^+) \,,\\
			&\mathbf{B}=\left(\begin{array}{ccc}
				\frac{1}{\sqrt{6}}\Lambda+\frac{1}{\sqrt{2}}\Sigma^{0} & \Sigma^{+} & p \\
				\Sigma^{-} & \frac{1}{\sqrt{6}}\Lambda-\frac{1}{\sqrt{2}}\Sigma^{0} & n \\
				\Xi^{-} & \Xi^{0} & -\sqrt{\frac{2}{3}}\Lambda
			\end{array}\right).
		\end{aligned}
	\end{equation}
The coefficients $a_n^f$ are $SU(3)_F$ singlets and, according to the generalized Wigner-Eckart theorem, can be parameterized as~\cite{footnote}
\begin{equation}\label{paras}
	\begin{aligned}
		a_{n}^{f} =& C_{n}^{f}(\mathbf{B}^{\dagger})^{\alpha}_{\beta}\mathcal{J}^{\beta}(\mathbf{B}_c)_{\alpha}+V_{n}^{f}(\mathbf{B}^{\dagger})^{\alpha}_{\gamma}\mathcal{J}^{\beta}\mathcal{S}^{\gamma}_{\beta}(\mathbf{B}_c)_{\alpha}\\
		&+W_{n}^{f} (\mathbf{B}^{\dagger})^{\gamma}_{\beta}\mathcal{J}^{\beta}\mathcal{S}^{\alpha}_{\gamma}(\mathbf{B}_c)_{\alpha}.
	\end{aligned}
\end{equation}
Here, $\mathcal{J}$ transforms in the same $SU(3)_F$ representation as $q $ in $c\rightarrow q  $ transitions.  The first-order $SU(3)_F$-breaking effects are captured by the insertions of ${\cal S}$ = diag$(1,1,-2)/\sqrt{6}$ with the parameters  
$V_{n}^{f}$ and $W_{n}^{f}$, which are of order $m_s/\Lambda_{\text{QCD}}$.  
In the $SU(3)_F$ limit,  $V_{n}^{f} = W_{n}^{f} =0 .$
In the above equations, the relevant components of $\mathbf{B}_c$ and $\mathbf{B}$ are selected by flavor indices. For instance, for the $c\to s $ transition of $\Lambda_c^+\to \Lambda$, one takes
\begin{equation}
\!\!\!\!	\mathbf{B}_c={\cal J}^\dagger  =(0,0,1),\quad
	\mathbf{B}=
	\mathrm{diag} 
	\left(\sqrt{\frac{1}{6}},\sqrt{\frac{1}{6}},-\sqrt{\frac{2}{3}}\right),
\end{equation} 
resulting in 
\begin{equation}\label{henew}
a_n^f ( \Lambda_c \to \Lambda) = -\sqrt{\frac23}C_n ^f + \frac23  V_n ^f+ \frac23  W_n ^f\,. 
\end{equation}
The same flavor structure applies to the other form factors as well. 

\begin{table}[ht]
	\renewcommand{\arraystretch}{2}
	\centering
	\begin{tabular}{ c|c|c|c }
		\hline
		$(f,n)$ &  $C^f_n$ &  $V^f_{n}$ & $W^f_{n}$   \\ 
		\hline
	  \makecell{$(f_{\perp},0)$ \\ $(f_{\perp},1)$ \\ $(f_{\perp},2)$ \\ $(f_{\perp},3)$} & \makecell{$1.528(71)$\\$-4.093(945)$\\$4.5(8.6)$\\$1(30)$} & \makecell{$-0.202(103)$\\$1.764(1.564)$\\$-8(14)$\\$7(38)$} & \makecell{$0.155(82)$\\$-2.502(1.436)$\\$3(13)$\\$-4(39)$} \\
		\hline
		\makecell{$(f_{+},0)$ \\ $(f_{+},1)$ \\ $(f_{+},2)$ \\ $(f_{+},3)$} & \makecell{$0.849(39)$\\$-3.310(679)$\\$11.7(6.8)$\\$-6(30)$} & \makecell{$-0.137(55)$\\$0.796(913)$\\$-2.3(8.9)$\\$9(38)$} & \makecell{$-0.009(40)$\\$-1.280(714)$\\$6.6(7.4)$\\$-16(38)$} \\
		\hline
		\makecell{$(f_{0},0)$ \\ $(f_{0},1)$ \\ $(f_{0},3)$} & \makecell{$0.838(40)$\\$-3.206(694)$\\$3(30)$} & \makecell{$-0.090(54)$\\$-0.206(924)$\\$8(38)$} & \makecell{$-0.023(36)$\\$-1.066(679)$\\$-4(38)$} \\
		\hline
		\makecell{$(g_{\perp},0)$ \\ $(g_{\perp},1)$ \\ $(g_{\perp},2)$ \\ $(g_{\perp},3)$} & \makecell{$0.689(26)$\\$-2.134(500)$\\$8.8(6.2)$\\$-8(28)$} & \makecell{$-0.115(36)$\\$1.077(722)$\\$-2.9(8.7)$\\$-17(36)$} & \makecell{$-0.046(26)$\\$-0.690(609)$\\$0.7(7.7)$\\$19(38)$} \\
		\hline
		\makecell{$(g_{+},0)$ \\ $(g_{+},1)$ \\ $(g_{+},2)$ \\ $(g_{+},3)$} & \makecell{$0.689(26)$\\$-2.161(434)$\\$5.8(5.4)$\\$-3(28)$} & \makecell{$-0.115(36)$\\$1.460(571)$\\$-11.8(6.9)$\\$-15(35)$} & \makecell{$-0.046(26)$\\$-0.251(413)$\\$-6.6(5.6)$\\$-3(37)$} \\
		\hline
		\makecell{$(g_{0},0)$ \\ $(g_{0},1)$ \\ $(g_{0},3)$} & \makecell{$0.737(40)$\\$-2.893(611)$\\$-2(29)$} & \makecell{$-0.140(56)$\\$2.277(821)$\\$2(37)$} & \makecell{$-0.037(39)$\\$-0.959(615)$\\$8(38)$} \\
		\hline
	\end{tabular}
	\caption{Numerical results for the BCL $z$ expansion with $SU(3)_F$ symmetry including first-order breaking. The uncertainties are taken from the LQCD calculations~\cite{Meinel:2016dqj,Meinel:2017ggx,Farrell:2025gis}. }
	\label{tab:su(3) paras}
\end{table}

\section{Numerical Results}
\label{sec:Numerical results}
We take the LQCD results~\cite{Meinel:2016dqj,Meinel:2017ggx,Farrell:2025gis} as input,   including the   form factors for $\Lambda_c^+\to\Lambda$, $\Lambda_c^+\to n $, and $\Xi_c\to\Xi$. From Eq.~\eqref{paras}
we   solve   $C_n^f , V_n^f $ and $W_n^f$ by 
\begin{equation}
\begin{pmatrix}
C_n^f \\[2em]
V_n^f \\[2em]
W_n^f
\end{pmatrix}
=
\begin{pmatrix}
\dfrac{1}{\sqrt{6}} & \dfrac{2}{3} & \dfrac{2}{3} \\[0.8em]
1 & \sqrt{\dfrac{2}{3}} & 0 \\[0.8em]
1 & 0 & \sqrt{\dfrac{2}{3}}
\end{pmatrix}
\begin{pmatrix}
a_n^f(\Lambda_c \to \Lambda) \\[2em]
a_n^f(\Lambda_c \to n) \\[2em]
a_n^f(\Xi_c \to \Xi)
\end{pmatrix}
\, .
\end{equation}
A slightly different BCL $z$-expansion was adopted in Refs.~\cite{Meinel:2016dqj,Meinel:2017ggx}. The corresponding expansion variable is similar to $z_f(q^2)$ defined in Eq.~\eqref{eq1}, except that $t_+^f = (m_D + m_{\pi(K)})^2$ for $f = g_0, g_+, g_\perp$. We denote the variable therein by $\tilde z_f(q^2)$ and match it to $z_f(q^2)$ used in this work, as described in Appendix~\ref{appendix}. 
The numerical results are presented in Table~\ref{tab:su(3) paras}. The resulting form factors for $\Xi_c\to \Xi$ are essentially identical to those from LQCD~\cite{Farrell:2025gis}, while we have checked that those for $\Lambda_c^+ \to \Lambda, n$ also reproduce the LQCD results up to ${\cal O}(10^{-3})$~\cite{Meinel:2016dqj,Meinel:2017ggx}.
Since $|z_f|<0.16$ over the physical region, the form factors are dominated by the terms with $n=0$. From the table, one can see that the typical sizes of $V^f_{0}/C^f_0$ and $W^f_{0}/C^f_0$ are around $10\%$ to $20\%$, as expected in the usual $SU(3)_F$ framework. 
Errors from truncating the $SU(3)_F$ breaking are therefore expected to be of order $10^{-2}$.

\begin{figure}[ht]
\centering 
\includegraphics[width=0.23\textwidth]{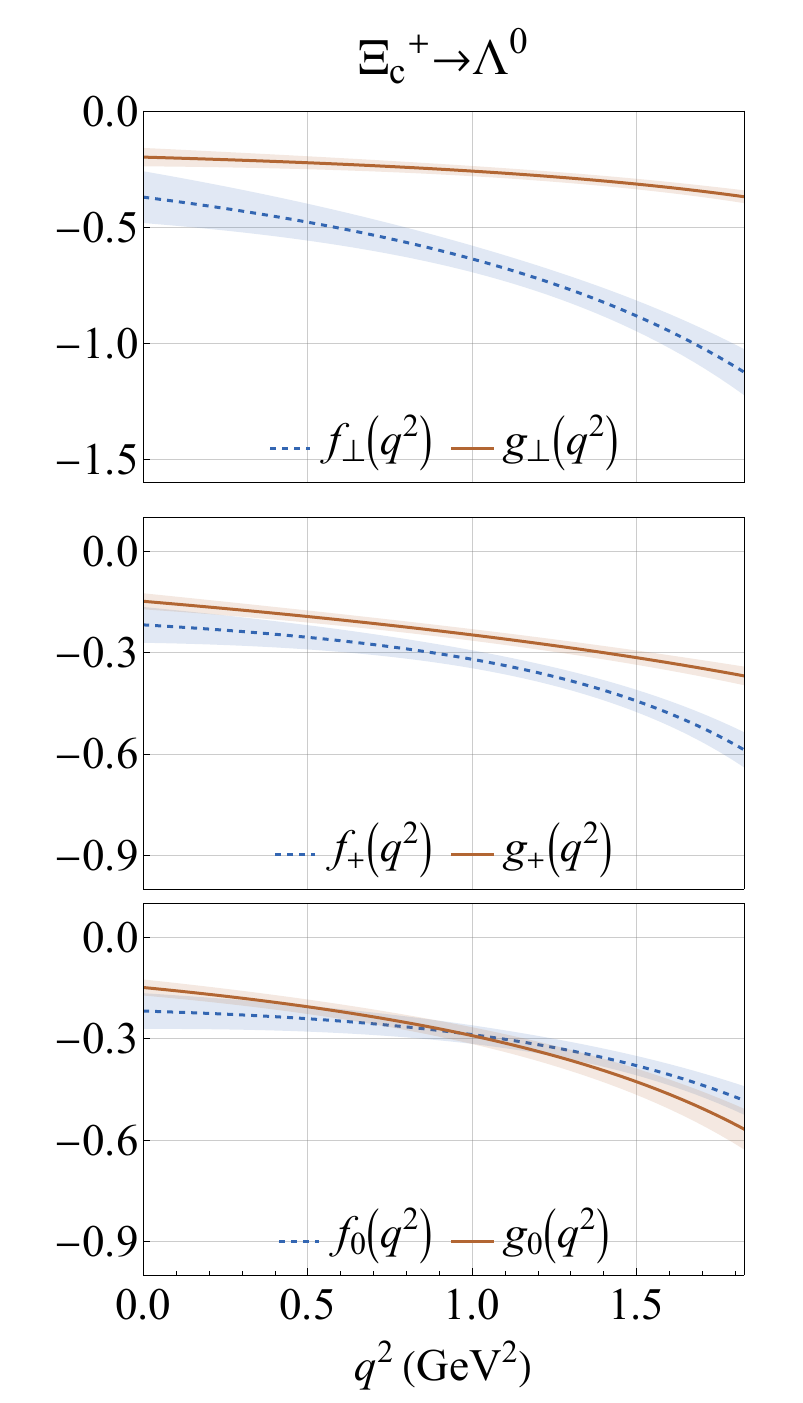}
\includegraphics[width=0.23\textwidth]{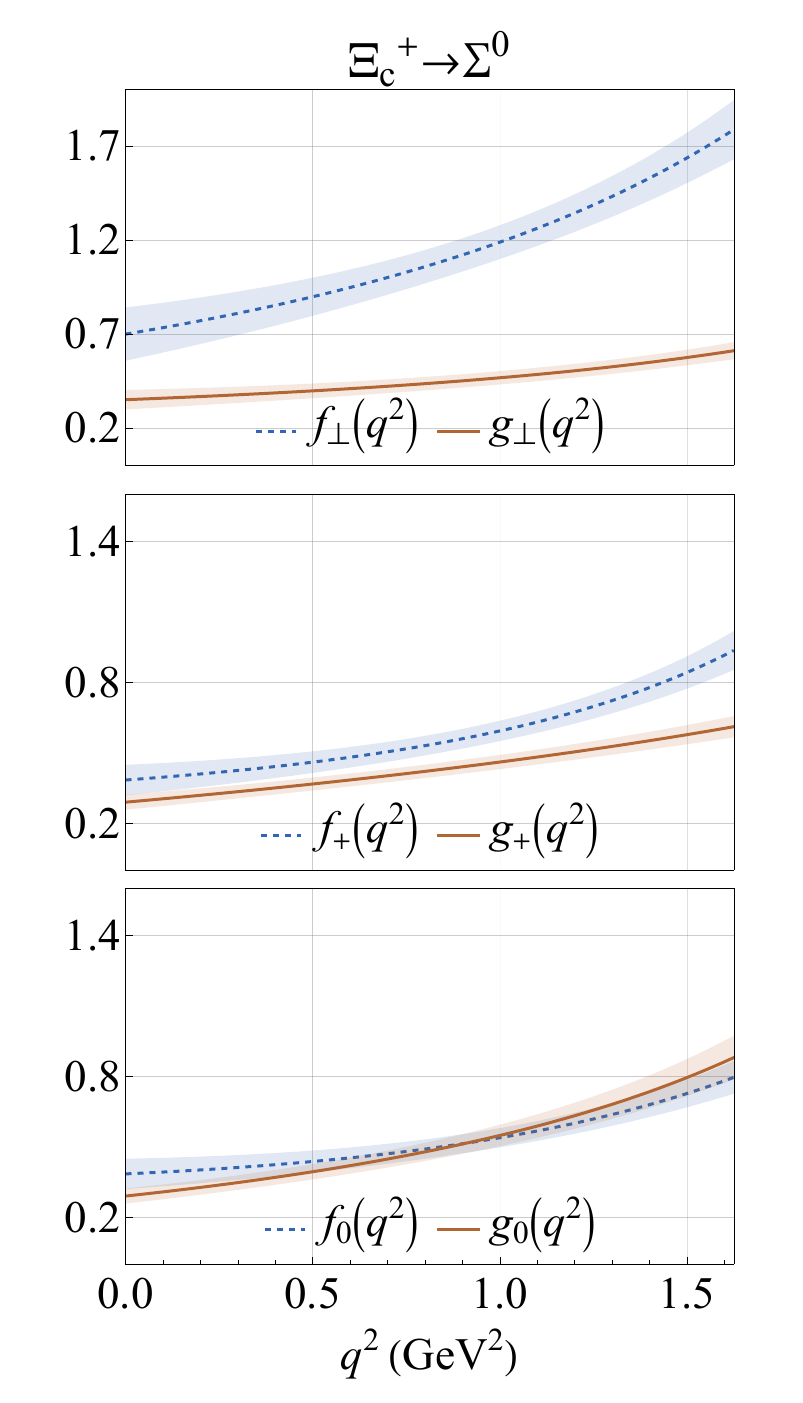}
\hfill
\caption{The numerical results of form factors $f_{\perp}$, $f_{+}$, $f_{0}$, $g_{\perp}$, $g_{+}$ and $g_{0}$ from the $SU(3)_F$ approach of $\Xi_c^+\rightarrow \Lambda$ and $\Xi_c^+\rightarrow \Sigma^0$.}
\label{fig:form factors}
\end{figure}

The genuinely new predictions are those for $\Xi_c^+ \to \Lambda $ and $\Xi_c^+ \to \Sigma^0 $ given in Table~\ref{tab:z-expansion-paras}, with the coefficients 
\begin{equation}
\!\!\!\!\begin{pmatrix}
a_n^f(\Xi_c^+  \to \Lambda) \\[2em]
a_n^f(\Xi_c^+ \to \Sigma ^0 ) 
\end{pmatrix}
=
\begin{pmatrix}
-\dfrac{1}{\sqrt6} & -\dfrac{1}{6} & -\dfrac{1}{6} \\[1.0em]
\dfrac{1}{\sqrt2} & \dfrac{1}{2\sqrt3} & \dfrac{1}{2\sqrt3}
\end{pmatrix}
\begin{pmatrix}
C_n^f \\[1em]
V_n^f \\[1em]
W_n^f
\end{pmatrix}
\, .
\end{equation}
The isospin symmetry implies that 
$a_n^f$ of 
$\Xi_c^0 \to \Sigma^-$ is $\sqrt{2}$ times larger than those of 
$\Xi_c^+ \to \Sigma^0$ and for simplicity we list only the form factors for the latter channel. Since $|z_f(q^2)| < 0.16$ for all form factors $f$, the contributions from higher-order terms are strongly suppressed. Consequently, the uncertainties of the form factors are dominated by those of the coefficients $a_n^f\,(C_n^f,V_n^f,W_n^f)$ with $n=0$ and $n=1$.
The form factors are explicitly depicted in 
Fig.~\ref{fig:form factors}. 

\begin{table}[ht]
	\renewcommand{\arraystretch}{1.2}
	\centering
\hspace{-0.28cm}	\begin{tabular}{ c|cc|cc }
		\hline
		\multirow{2}{*}{$(i,n)$} 
		& \multicolumn{2}{c|}{$\Xi_c^+\rightarrow\Lambda$} 
		& \multicolumn{2}{c}{$\Xi_c^+\rightarrow\Sigma^0$} \\
		\cline{2-3}\cline{4-5}
		& $a^{f_i}_{n}$ & $a^{g_i}_{n}$ & $a^{f_i}_{n}$ & $a^{g_i}_{n}$ \\
		\hline
		$(\perp,0)$ & $-0.62(5)$      & $-0.25(2)$      & $1.07(9)$       & $0.44(3)$ \\
		$(\perp,1)$ & $1.79(82)$      & $0.81(39)$      & $-3.11(1.42)$    & $-1.40(67)$ \\
		$(\perp,2)$ & $-1(7)$    & $-3.2(4.8)$    & $2(13)$   & $5.6(8.3)$ \\
		$(\perp,3)$ & $-1(23)$  & $3(22)$   & $1(39)$   & $-5(38)$ \\
        \hline
		$(+,0)$     & $-0.32(3)$      & $-0.25(2)$      & $0.56(5)$       & $0.44(3)$ \\
		$(+,1)$     & $1.43(49)$      & $0.68(31)$      & $-2.48(85)$     & $-1.18(53)$ \\
		$(+,2)$     & $-5(5)$    & $0.7(3.8)$     & $9.5(8.6)$     & $-1.2(6.7)$ \\
		$(+,3)$     & $4(22)$   & $4(21)$   & $-6(39)$  & $-8(37)$ \\
               \hline
		$(0,0)$     & $-0.32(3)$      & $-0.27(3)$      & $0.56(5)$       & $0.47(5)$ \\
		$(0,1)$     & $1.52(49)$      & $0.96(44)$      & $-2.64(86)$     & $-1.67(76)$ \\
		$(0,3)$     & $-2(22)$  & $-1(22)$  & $3(39)$   & $2(38)$ \\
		\hline
	\end{tabular}
	\caption{The numerical results of  $z$-expansion parameters for  $\Xi_c^+\rightarrow\Lambda$ and $\Xi_c^+\rightarrow\Sigma^0$. 
    The parameters of $a_2^{f_0} $ and 
     $a_2^{g_0} $ can be obtained by Eq.~\eqref{match}. 
    }
	\label{tab:z-expansion-paras}
\end{table}

In addition to the branching fractions, experiments have measured the up-down asymmetries defined by 
\begin{equation}
\alpha=\frac{\Gamma_{+}-\Gamma_{-}}{\Gamma_{+}+\Gamma_{-}} \,,
\end{equation}
where $\Gamma_{\pm}$ denote the partial rates into final-baryon helicity $\lambda_{\mathbf{B}}=\pm1/2$.
For $\Lambda_c^+ \to\Lambda e ^+ \nu_e $, it can be measured by the sequential decay of $\Lambda \to p \pi ^- $~\cite{BESIII:2019odb,Belle:2022uod,BESIII:2022udq,BESIII:2023jxv,LHCb:2024tnq}.
Utilizing the CKM matrix elements $V_{cd}=0.225\pm0.004$ and 
$V_{cs}=0.975\pm0.006$ from the latest CKMfitter results~\cite{Charles:2004jd}, 
together with the lifetimes of singly charmed baryons reported by the PDG~\cite{ParticleDataGroup:2024cfk}, we evaluate the branching fractions and up-down asymmetries for channels $\Xi_c^+\rightarrow \Lambda$ and $\Xi_c\rightarrow \Sigma$ in Table~\ref{tab:br and upa}, where the results are genuine new predictions established in this work.

\begin{table}[ht]
	\renewcommand{\arraystretch}{2}
	\centering
	\begin{tabular}{c|c|c}
		\hline
		Channels  & $\mathcal{B}(SU(3)_F)\,(\%)$ & $\alpha(SU(3)_F)\,(\%)$ \\
				\hline
		\makecell[l]{$\Xi_c^+\rightarrow \Sigma^0e^+\nu_e$\\$\Xi_c^+\rightarrow \Sigma^0\mu^+\nu_\mu$}
		& \makecell[c]{$0.282(30)$\\$0.275(29)$}
		& \makecell[c]{$-89.9(9.3)$\\$-89.7(9.2)$} \\
        \hline
        \makecell[l]{$\Xi_c^+\rightarrow \Lambda e^+\nu_e$\\$\Xi_c^+\rightarrow \Lambda \mu^+\nu_\mu$}
		& \makecell[c]{$0.117(14)$\\$0.115(13)$}
		& \makecell[c]{$-90.8(10.2)$\\$-90.7(10.1)$} \\
		\hline
		\makecell[l]{$\Xi_c^0\rightarrow \Sigma^-e^+\nu_e$\\$\Xi_c^0\rightarrow \Sigma^-\mu^+\nu_\mu$}
		& \makecell[c]{$0.186(20)$\\$0.182(19)$}
		& \makecell[c]{$-89.8(9.3)$\\$-89.7(9.2)$} \\
        \hline
	\end{tabular}
	\caption{Numerical results for branching fractions and up-down asymmetries in the $SU(3)_F$ approach for the channels $\Xi_c^+\rightarrow \Lambda$ and $\Xi_c\rightarrow \Sigma$, in units of percentage.}
	\label{tab:br and upa}
\end{table} 

As  discussed in the introduction, the puzzle 
may be traced back to the normalization channel. To test the $SU(3)_F$ breaking in a normalization-independent way, we propose to 
measure the relative  branching fractions to  $\Xi_c \rightarrow\Xi \ell^+\nu_{\ell}$. In particular, we predict  the ratios 
\begin{equation}\label{br_ratio}
    \begin{aligned}
R_{\Sigma ^0}  & \equiv 
\frac{\mathcal{B} (\Xi_c^+\to\Sigma^0\ell^+\nu_{\ell})}{\mathcal{B}(\Xi_c^+\to\Xi^0\ell^+\nu_{\ell})}  =( 2.6\pm0.3)\% , \\
R_{\Lambda }& \equiv 
\frac{\mathcal{B}(\Xi_c^+\to\Lambda \ell^+\nu_{\ell})}{\mathcal{B}(\Xi_c^+\to\Xi^0\ell^+\nu_{\ell})} =( 1.1\pm0.1)\%, \\ 
    R_{\Sigma ^- } & \equiv  \frac{\mathcal{B}(\Xi_c^0\to\Sigma^-\ell^+\nu_{\ell})}{\mathcal{B}(\Xi_c^0\to\Xi^-\ell^+\nu_{\ell})}  = 
    2 R_{\Sigma^0}
 ,
    \end{aligned}
\end{equation}
which are protected up to second order in $SU(3)_F$ breaking, with expected precision at the percent level.
These ratios have the advantage of avoiding the normalization ambiguity associated with $\mathcal{B}(\Xi_c^0\rightarrow\Xi^-\pi^+)$.

\begin{table}[h]
	\centering
	\renewcommand{\arraystretch}{1.2}
	\begin{tabular}{l|c cc}
		\hline
		Values & Exact $SU(3)_F$ & $SU(3)_F$+LQCD
        &$
        SU(3)_F
        $+ Exp.~\cite{He:2021qnc}\\ 
		\hline
		$R_{\Sigma^0} (\%)$ & 2.6 &$2.6\pm0.3$ & $12.8 \pm 2.7$ \\
		$R_{\Lambda} (\%)$ & 0.9  &$1.1\pm0.1$  & $1.7 \pm 0.5$ \\ 
		\hline
	\end{tabular}
	\caption{Values of $R_{\Sigma^0}$ and $R_{\Lambda}$ for the exact $SU(3)_F$ symmetry scheme, for our $SU(3)_F$ symmetry analysis matched to LQCD inputs from Eq.~\eqref{br_ratio}, and for the first-order
		$SU(3)_F$-breaking analysis of Ref.~\cite{He:2021qnc}, where the input
		parameters were determined from experimental data. 
    }
	\label{tab:su3+exp}
\end{table}

In Table~\ref{tab:su3+exp}, we compare the values of $R_{\bf B}$ obtained in the exact $SU(3)_F$ limit, in our $SU(3)_F+\mathrm{LQCD}$ analysis, and in the first-order $SU(3)_F$-breaking analysis using the experimental inputs
${\cal B}(\Lambda_c^+ \to \Lambda \ell^+ \nu_\ell)$ and
${\cal B}(\Xi_c \to \Xi \ell^+ \nu_\ell)$~\cite{He:2021qnc}. 
Our results are very close to the exact-$SU(3)_F$ predictions, indicating that the relevant first-order $SU(3)_F$-breaking effects are small according to LQCD.  
By contrast, the data-driven first-order analysis~\cite{He:2021qnc} gives a much larger value of $R_{\Sigma^0}$, about five times larger than both the exact-$SU(3)_F$ and $SU(3)_F+\mathrm{LQCD}$ predictions.  
If future measurements favor the exact-$SU(3)_F$ and $SU(3)_F+\mathrm{LQCD}$ predictions, this would point to the normalization input used in Ref.~\cite{He:2021qnc},
$ 
{\cal B}(\Xi_c^0 \to \Xi^- \pi^+) = (1.80\pm0.50\pm0.14)\%
$ 
as a likely source of the tension.  
In this case, since the semileptonic ratios would be consistent with both $SU(3)_F$ symmetry and LQCD, the discrepancy would more naturally be attributed to the experimental normalization channel $\Xi_c^0\to\Xi^-\pi^+$ than to large unknown hadronic effects in the semileptonic form factors.
Conversely, if the measurements favor the data-driven first-order result, this would suggest that additional hadronic effects beyond those captured by the LQCD-matched $SU(3)_F$ analysis may be required. 

We now discuss the experimental testability. The numbers of observed events are expected to be 
\begin{equation}
\frac{
N( \Xi_c \to {\bf B} e^+ \nu_e )
}{N( \Xi_c  \to \Xi  e^+ \nu_e )}
= R_{{\bf B}} 
\frac{\epsilon_\mathbf{B}}{
\epsilon_{\Xi }
}\,,
\end{equation}
where $\epsilon_{\bf B} $ is the reconstruction efficiency for the final-state baryon ${\bf B}$. 
The number of produced $\Xi_c^+$ baryons at Belle and Belle~II is not
directly known. For an order-of-magnitude estimate, we assume comparable
production rates for $\Xi_c^+$ and $\Xi_c^0$, as suggested by approximate
isospin symmetry in charm fragmentation.
This approximation is based on the assumption that the $c\bar c$ pair is produced via $\gamma^*$, thereby minimizing the number of hard gluons, along with the use of isospin symmetry.
We take the efficiencies  to 
be 
$(\epsilon_{\Xi^0} , 
\epsilon_{\Sigma ^0}, 
\epsilon_{\Lambda },
\epsilon_{\Xi^- }
)  = (1.5\%,7\%,20\%,20\%) $~\cite{Belle:2024ikp,Belle:2021zsy,Belle:2021avh}.
In 2021, $(16.01\pm0.44)\times10^3$ events were observed for $\Xi_c^0 \to \Xi^- e^+ \nu_e$ with an integrated luminosity of $711+89.5=800.5~\mathrm{fb}^{-1}$ at $\sqrt{s}=10.58$ and $10.52~\mathrm{GeV}$ at Belle~\cite{Belle:2021crz}. 
At present, the combined integrated luminosity of Belle and Belle~II has
reached $1.8\,{\rm ab}^{-1}$~\cite{Belle:2017ext,BelleII:homepage}.
Based on the Belle measurements and naively assuming the same reconstruction
efficiencies at Belle and Belle~II, we estimate the expected statistical
precisions to be
$(\delta R_{\Sigma^0},\delta R_{\Lambda}) \approx (0.40\%,0.13\%) .$
This estimate should be regarded as conservative, since the improved
performance of the Belle~II detector is expected to lead to higher
reconstruction efficiencies than those at Belle
~\cite{Aihara:2024zds,Belle-II:2025wpi,Belle-II:2025tpe}. With the future
Belle~II data set, whose integrated luminosity is expected to reach
$50~\mathrm{ab}^{-1}$~\cite{Belle-II:2018jsg}, the statistical precision
would improve by at least about a factor of six.

\begin{figure}[ht]
\centering 
\includegraphics[width=0.5\textwidth]{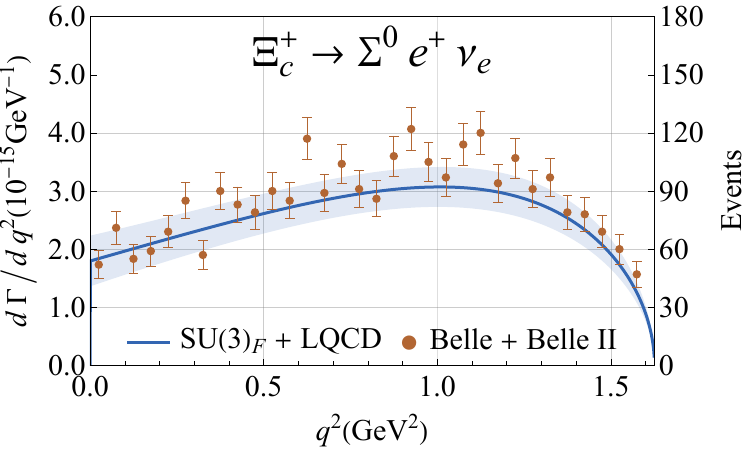}
\includegraphics[width=0.5\textwidth]{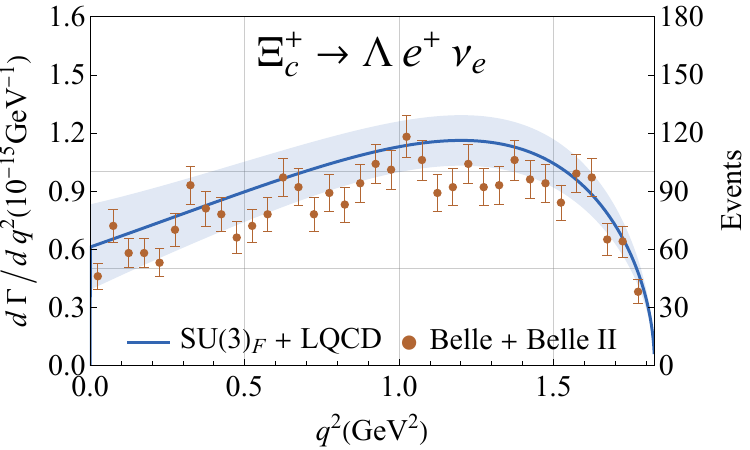}
\caption{Differential decay rates and estimated event yields for
	$\Xi_c^+ \to \Sigma^0 e^+ \nu_e$ and
	$\Xi_c^+ \to \Lambda e^+ \nu_e$. The $q^2$ dependence of the differential
	decay rates is shown on the left $Y$-axis, while the event yields estimated
	by scaling to $\mathcal L_{\rm int}=1.8~{\rm ab}^{-1}$ are shown on the
	right $Y$-axis.}
\label{fig:differential decay rates}
\end{figure}

The simulations of the experiments for these channels are presented in Fig.~\ref{fig:differential decay rates}, and 
the label ``Belle + Belle~II'' denotes an order-of-magnitude scaling to a combined integrated luminosity $\mathcal{L}_{\rm int}=1.8~\mathrm{ab}^{-1}$.
Throughout the simulations, we assume that the detection efficiencies are independent of $q^2$, while the $\sqrt{q^2}$-bin widths are fixed at $50$~MeV. From the figure, one sees that the expected signal yields are sizable across a wide range of $\sqrt{q^2}$, suggesting that the differential spectra are experimentally measurable with sufficient statistics.

Since the dominant decay $\Sigma^- \to n\pi^-$ contains a neutron in the
final state, this channel is difficult to reconstruct at Belle and Belle~II.
It could, however, be reconstructed at BESIII and STCF, with an estimated
efficiency $\epsilon_{\Sigma^-}=14.3\%$~\cite{BESIII:2025kna}. So far, the production numbers of $\Xi_c$ are not known. Assuming that the near-threshold $\Xi_c\overline{\Xi}_c$ production rate is about $0. 1\sim 0.3  $ of that for $e^+e^- \to \Lambda_c^+\bar{\Lambda}_c^-$, one is led to a benchmark cross section
$\sigma(e^+e^- \to \Xi_c\bar{\Xi}_c)\sim 10\text{--}30~\mathrm{pb}$~\cite{BESIII:2023rwv, BESIII:2022ulv}.
With the approved BEPCII upgrade, the maximum center-of-mass energy will be extended to $5.6~\mathrm{GeV}$, while the peak luminosity at high energies is expected to increase by about a factor of three~\cite{BESIII:2022mxl}. 
This will enable direct studies of $\Xi_c\bar{\Xi}_c$ production; for a benchmark cross section $\sigma(e^+e^-\to\Xi_c\bar{\Xi}_c)\sim10$--$30~\mathrm{pb}$, one expects roughly $(1$--$3)\times10^5$ produced pairs per year at peak luminosity at BEPCII-U, while the corresponding yield at the STCF would be about $(1$--$3)\times10^7$ pairs per year~\cite{Achasov:2023gey}. 

\section{Conclusions}\label{sec:Conclusions}
We have developed an $SU(3)_F$ analysis of singly charmed baryon semileptonic decays by matching the available lattice-QCD form factors onto the BCL $z$-expansion, including first-order symmetry-breaking effects. The resulting framework consistently reproduces the known lattice inputs and yields symmetry-breaking corrections of the expected size.

With these inputs, we have predicted the form factors and decay observables for $\Xi_c^+\to \Lambda \ell^+\nu_\ell$ and $\Xi_c\to \Sigma \ell^+\nu_\ell$. In particular, we identified the ratios
$ 
(R_{\Sigma^-},R_{\Sigma^0},R_{\Lambda})=(5.2\pm0.6,\;2.6\pm0.3,\;1.1\pm0.1)\% \, .
$
These ratios are free from the normalization ambiguity associated with ${\cal B}(\Xi_c^0\to \Xi^-\pi^+)$ and ${\cal B}(\Xi_c^+\to \Xi^-\pi^+\pi^+)$. They therefore provide normalization-independent probes of the first-order $SU(3)_F$ breaking.

We have also shown that the corresponding signal yields should be
experimentally measurable within the  Belle + Belle~II luminosity
estimate, subject to detector-specific efficiencies and
systematic uncertainties. The expected statistical precisions are
$(\delta R_{\Sigma^0},\delta R_{\Lambda})
\simeq (4.0,\,1.3)\times 10^{-3}.$ Precise determinations of these ratios would provide normalization-independent
tests of $SU(3)_F$ breaking and help clarify whether the current
puzzle originates from an underestimation of the nonleptonic $\Xi_c^0\to\Xi^-\pi^+$ decay or from unknown hadronic effects.

\appendix

\section{Matching $z$-expansion conventions} 
\label{appendix}

Expanding $\tilde z_f $ in powers of $z_f$, we obtain formally 
\begin{equation}
\tilde z_f  =   d_0 + d_1 z_f + d_2 z_f^2 + d_3 z_f^3 
+ {\cal O}(z_f^4)    \, ,
\end{equation}
where $d_n$ is defined by 
\begin{equation}
d_n^f
=
\frac{1}{n!}
\left.
\frac{d^n \tilde z_f}{d z_f^n}
\right|_{z_f=0}
=
\frac{1}{n!}
\left.
\left(
\frac{dq^2}{dz_f}\frac{d}{dq^2}
\right)^n
\tilde z_f(q^2)
\right|_{z_f=0}.
\end{equation}
The explicit form is too lengthy to be informative and can be calculated by a modern computer program.

We denote the BCL coefficients in the expansion in $\tilde z_f$ by $\tilde a_n^f$, and impose the matching condition
\begin{equation}
\sum_n a_n^f z_f^n
=
\sum_n \tilde a_n^f \tilde z_f^n \, .
\end{equation}
This leads to the relations
\begin{equation}
\begin{aligned}
a_0^f =&\,\, \tilde a_0^f + \tilde a_1^f d_0^f + \tilde a_2^f (d_0^f)^2 + \tilde a_3^f (d_0^f)^3 \,,\\
a_1^f =&\,\, \tilde a_1^f d_1^f + 2 \tilde a_2^f d_0^f d_1^f + 3 \tilde a_3^f (d_0^f)^2 d_1^f \,,\\
a_2^f =&\,\, \tilde a_1^f d_2^f + \tilde a_2^f \!\left[ (d_1^f)^2 + 2 d_0^f d_2^f \right]\\&\,\,
+ \tilde a_3^f \!\left[ 3 d_0^f (d_1^f)^2 + 3 (d_0^f)^2 d_2^f \right] \,,\\
a_3^f =&\,\, \tilde a_1^f d_3^f + 2 \tilde a_2^f \!\left( d_0^f d_3^f + d_1^f d_2^f \right)
\\&\,\, + \tilde a_3^f \!\left[ (d_1^f)^3 + 6 d_0^f d_1^f d_2^f + 3 (d_0^f)^2 d_3^f \right] \, ,\end{aligned}
\end{equation}
up to the precision  of $|z_f^4(q^2)| <  10 ^{-3}. $
We note that $\tilde a_n^f$ is denoted simply by $a_n^f$ in Refs.~\cite{Meinel:2016dqj,Meinel:2017ggx}. The above relations therefore allow us to match their parametrization to the one in this work.   

\vspace{1em}
\acknowledgments

We would like to express our sincere appreciation to Long-Ke Li for the valuable discussions. This work is supported in part by the National Natural Science Foundation of China (NSFC) under Grant No. 12547104 and 12575096.

\end{document}